\documentclass[journal = jctcce]{achemso}
\usepackage{amsmath}
\usepackage{amssymb}
\usepackage{dcolumn} 
\usepackage{setspace}
\usepackage{array}
\usepackage{graphicx}
\usepackage{booktabs}
\usepackage{multirow}
\usepackage{psfrag}
\nocite{*} 

\def\gtg{{\footnotesize{$
\begin{array}{c}
{\rm G} \\
{\rm C}
\end{array}
\left(
\begin{array}{c}
{\rm T} \\
{\rm A}
\end{array}
\right)_N
\begin{array}{c}
{\rm G} \\
{\rm C}
\end{array}
$}}}
\def\atn{{\footnotesize{$
\left(
\begin{array}{c}
{\rm T} \\
{\rm A}
\end{array}
\right)_N
$}}}

\def\sgtg{{G(T)$_N$G}}
\def\sgag{{G(A)$_N$G}}

\def\brp{{\mathbf{r}^{\prime}}}

\def\br{{\mathbf{r}}}

\def\bR{{\mathbf{R}}}
\def\bS{{\mathbf{S}}}

\def\d{{\mathrm{d}}}
\def\rhor{{\rho({\bf r})}}

\def\rhoi{{\rho_I}}

\def\rhoir{{\rho_I({\bf r})}}

\def\rhojrp{{\rho_J({\bf r}^{\prime})}}
\def\sumi{{\sum_I^{N_S}}}

\newcommand{\michele}[1]{{#1}}

\newcommand{\eqn}[1]{\ref{#1}}
\newcommand{\eqs}[2]{\ref{#1}--\ref{#2}}
\newcommand{\pot}[1]{v_{\rm #1}}

\author{Pablo Ramos}
\author{Michele Pavanello}
\email{m.pavanello@rutgers.edu}
\affiliation{Department of Chemistry, Rutgers University, Newark, NJ 07102, USA}
\title[Hole Transfer Couplings in Biosystems]{Quantifying Environmental Effects on the Decay of Hole Transfer Couplings in Biosystems}

\begin{document}
\maketitle
\section*{Abstract}
In the past two decades, many research groups worldwide have tried to understand and categorize simple regimes in the charge transfer of such biological systems as DNA. Theoretically speaking, the lack of exact theories for electron--nuclear dynamics on one side, and poor quality of the parameters needed by model Hamiltonians and nonadiabatic dynamics alike (such as couplings and site energies) on the other, are the two main difficulties for an appropriate description of the charge transfer phenomena. In this work, we present an application of a previously benchmarked and linear-scaling subsystem DFT method for the calculation of couplings, site energies and superexchange decay factors ($\beta$) of several biological donor--acceptor dyads, as well as double stranded DNA oligomers comprised of up to 5 base pairs. The calculations are all-electron, and provide a clear view of the role of the environment on superexchange couplings in DNA -- they follow experimental trends and confirm previous semiempirical calculations. 
The subsystem DFT method is proven to be an excellent tool for long-range, bridge-mediated coupling and site energy calculations of embedded molecular systems. 
 \newpage
 \section{Introduction}
For the past two decades, charge transfer (CT) phenomena in biosystems have been intensively studied due to their role in biological functions as well as in potential applications related to nanosensors and molecular optoelectronics. Oxidative damage in cells \cite{gene2010,nunez1999,ratn1999} as well as the possibility of using DNA as a biomolecular nanowire \cite{gene2010} have inspired studies related to the hole migration through the DNA nucleobases on a sequence of radical cation states. 
A major target for oxidants is guanine (G), the nucleobase with the lowest ionization potential of the four DNA bases. Oxidation of guanine leads to a radical cation $G^{\bullet +}$, the hole may transfer to a neutral $G$, restoring neutrality in the former and creating a radical cation in the latter.
Especially the CT in DNA oligomers has been extensively studied both experimentally and theoretically with many techniques. 
Experiments have initially yielded different results, such as DNA being a conductor \cite{kasu2001}, semiconductor \cite{pora2000}, and insulator \cite{brau1998}. Also the theoretical understanding has been divergent and results in two points of view \cite{venk2011}. One claiming that DNA conduction occurs via a polaron picture by which the charge (e.g.\ a hole) is delocalized over a few nucleobases, and the resulting polaron hops along the DNA \cite{conw2005}. The other view is rooted in a conservative interpretation of the experiments by Giese {\em et al.} \cite{gies2000,gies2001} according to which charge transport over short distances occurs through superexchange-type conduction, while a long-range charge transfer is achieved by a combination of superexchange tunneling events as well as multistep hopping \cite{Lewis01102002,Lewis01081997,ja993689k}. In this interpretation, the excess charges are assumed to be localized on single nucleobases, and no polarons are invoked. Recently \cite{rena2013}, a hybrid 
mechanism that cannot be classified as hopping nor as coherent superexchange has been theoretically proposed for the long range CT in DNA. \michele{Incoherent mechanisms are also significant in DNA charge transport \cite{troi2003b,skou2010}. However, in this work we will only consider coherent transport. Decoherence can be taken into account theoretically by carrying out a simultaneous quantum dynamics of the electrons and the nuclei \cite{tully1998,land2011}, by ad-hoc corrections of the kinetic constant expression \cite{troi2003b,skou2010} or, at least partially, by accounting for the influence of a dissipative environment (phonon bath) which can support or restrict charge motion \cite{guti2009,kuba2013}.}

Theoretically, it has been difficult to provide a clear picture of DNA conductivity due to charge transfer dynamics being inherently non-adiabatic, and its modeling must involve going beyond the Born-Oppenheimer approximation. This was attempted using model Hamiltonians \cite{guti2009,groz1999,groz2000,groz2008a,senth2005,rena2013} as well as full electron-nuclear non-adiabatic dynamics simulations \cite{kuba2013,kubar2008a,kubar2009a}. However, the excess charge localization is dramatically affected by the specific theory employed for handling non-adiabaticity (e.g.\ Ehrenfest or Surface Hopping) \cite{kuba2013}.

Charge dynamics carried out with model Hamiltonians has been successful in elucidating certain regimes of CT in DNA. However, it has not offered a breakthrough. As an example, depending on the specific Hamiltonian used, the modeled long-range charge dynamics may \cite{guti2009} or may not \cite{groz2008a} agree with the experimental observations and generally only semiquantitative agreement with the experiments has been obtained.

It is generally recognized that the key to obtaining even semiquantitative agreement with the experiment is the accounting of structural fluctuations in the calculation of the parameters of the model Hamiltonians \cite{groz2002,troisi2002,groz2008,voit2008a,voit2008b}, as well as a proper accounting of the effect of the counterstrand on the nucleobases' ionization potentials \cite{sait1998,voit2000,voit2008b}. The effect of polarization by the molecular environment surrounding a nucleobase on its ionization potential has been reported to be up to 0.4 eV  \cite{voit2000}. Thus, accounting for these environmental effects is a major component of the modeling of charge transport in biological systems.

Current computational methods do not allow the fully quantum mechanical description of a biological molecule, such as solvated DNA, as the computational scaling of even mainstream Density-Functional Theory (DFT) methods goes roughly as $O(N^3)$, with $N$ being the number of atoms in the system. As the scale of model systems considered in theoretical simulations increases, the computational complexity quickly escalates to unreachable CPU times, even by the most powerful supercomputer on Earth. Thus, theoreticians are forced to make approximations (such as QM/MM treatments) to recover the environmental effects that are so important for the correct description of CT phenomena in biosystems. Approximate QM/MM methods are not ideal, as they need semiempirically determined parameters to function. In addition, even if the most accurate polarizable force field available is used \cite{leep2013} the wavefunction of the electrons belonging to the environment is completely lost, and with that typical quantum mechanical effects, 
such as electronic exchange and correlation interactions with the environment are generally not contemplated.

One of us has recently contributed to the development of an accurate and linear-scaling DFT method that allows to include environmental effects in the ionization potentials (site energies) as well as in the electronic Hamiltonian coupling matrix elements \cite{pava2013a,pava2011b} by exploiting ideas and techniques of subsystem DFT \cite{sen1986,cort1991,weso1993,hong2006}. We will refer to this method as Frozen Density Embedding (FDE) throughout this paper. This method has allowed us to approach system sizes that were simply not reachable before, especially for calculations of electronic couplings.

This work focuses on DNA and specifically on the effects of the molecular environment surrounding the nucleobases (particularly sugar groups and partner strand) involved in the hole transfer. \michele{Besides the solvating effects of the water molecules surrounding the DNA \cite{pava2010a, wolt2013}, the environment affects the hole transfer in DNA in several ways. For example, we will show that the sugar groups and the counterstrand stabilize the hole {\bf unevenly} when it is localized on different nucleobases (e.g.\ G is stabilized differently from T or A, also the 3$^\prime$ position is stabilized differently from 5$^\prime$). Counterions also play a role. Coupling of the hole motion with the motion of the counterions has been considered in the literature, and ion-gated charge transport has been proposed \cite{schu2004}. Because in this work we do not consider dynamical effects, we cannot infer on the effects of the counterions. We instead focus on completely desolvated DNA model systems employing an accurate and all-electron electronic structure method: the Frozen Density 
Embedding (FDE) formulation of subsystem DFT. In the most computationally demanding simulation, we calculated site energies and Hamiltonian couplings for the hole transfer in a DNA pentamer model system containing 308 atoms and 1322 interacting electrons. }

\michele{Experimental works report the so-called decay factor $\beta$ by fitting the charge transfer kinetics against the Marcus equation, which is derived from the Fermi golden rule electron transfer rate equation \cite{voit2002,marcus1985,dutt1992}, namely}
\begin{equation}
\label{marcus}
k_{CT} = \frac{2\pi}{\hbar} |V_{DA}|^2 FCWD.
\end{equation}
\michele{Where the Frank-Condon coefficient that within the Marcus theory is approximated as \cite{marcus1956,jort1976}:}
\begin{equation}
\label{fracon}
FCWD(\Delta E) = \frac{1}{\sqrt{4 \pi  E_r  k_B T}} \exp\left(-\frac{(\Delta E - E_r)^2}{4E_r k_B T}\right),
\end{equation}

\michele{where $\Delta E$ is the energy gap between the the initial and final state of electron transfer, $E_r$ is the reorganization energy, and $k_B$ is the Boltzman constant. The above equation is appropriate for the superexchange regime \cite{Nitzan_book}, i.e.\ when donor--acceptor energy levels are non-resonant to the bridge levels. Under those energetic conditions, the kinetic constant takes the following form, }
\begin{equation}
\label{m-l-j}
k_{CT}(R)=\frac{2\pi}{\hbar} \underbrace{V^2_0 \exp\left(-\beta R\right)}_{\left|V_{DA}\right|^2} FCWD, 
\end{equation}
where $R$ is an effective donor--acceptor separation distance.
Kinetic constants in the hopping regime, instead, follow a power law decay  \cite{bixo2001,reng2003,bixo1999,berl2005}.

\michele{In this work, we calculate effective donor--acceptor hole transfer couplings and fit them to \eqn{m-l-j}. We  first} focus on simple donor-acceptor systems, such as $\pi$-stacks dimers of nucleobases guanine ($G$), thymine ($T$), adenine ($A$). We also consider several other biologically relevant molecules, such as histidine, tyrosine, tryptophan, as well as other molecules rich in $\pi$ electrons that may be active in biological hole transfer. This first set of calculations will allow us to establish that FDE can calculate electronic couplings and site energies for biologically relevant dyads over a wide range of intermolecular separations. This is not a trivial achievement, as when the molecular fragments are separated by large distances, numerical inaccuracies can creep in and undermine the fidelity of the calculations. We propose effective solutions to such inaccuracies so that the calculated couplings are numerically stable over a wide range of intermolecular separations.
Once the simple donor-acceptor couplings are presented, we also treat superexchange in the B-DNA sequences \sgtg \ and \sgag \ with $N=$0-3, where the hole transfer is restricted to run on one strand. We analyzed the influence of the presence of the second strand and the sugar groups on the couplings, site energies and the overall decay factor, $\beta$, in the superexchange regime. 
 
\subsection{Frozen Density Embedding formulation of subsystem DFT}
Regular DFT is also known as Kohn--Sham DFT (KS-DFT), and  can be summarized by the following equation, the KS equation in canonical form,
\begin{equation}
\label{ks1}
\left[ -\frac{1}{2} \nabla^2 + \pot{eff}(\br) \right] \phi_k(\br) = \varepsilon_k  \phi_k(\br),
\end{equation}
where $\pot{eff}$ is the effective potential that the one-particle KS orbitals, $\phi_k$, experience, and $\varepsilon_k$ are the KS orbital energies. The spin labels have been omitted for sake of clarity. The electron density for singlets is simply $\rhor=2\sum_i^{\rm occ} \left| \phi_i(\br)\right|^2$.

The effective potential, $\pot{eff}$, is given by
\begin{equation}
\label{ks2}
\pot{eff}(\br)=\pot{eN}(\br)+\pot{Coul}(\br)+\pot{xc}(\br),
\end{equation}
where $\pot{eN}$ is the electron--nucleus attraction potential, $\pot{Coul}$ the Hartree potential, and $\pot{xc}$ the exchange--correlation (XC) potential\cite{kohn1965}. 

Subsystem DFT, instead, is based on the idea that a molecular system can be more easily approached if it is partitioned into many smaller subsystems. In mathematical terms, this is done by partitioning the electron density as follows \cite{sen1986,cort1991}
\begin{equation}
\label{fde1}
\rhor=\sumi \rhoir,
\end{equation}
with $N_S$ being the total number of subsystems.

Self-consistent solution of the following coupled KS-like equations (also called KS equations with constrained electron density \cite{weso2006}) yields the set of subsystem KS orbitals, i.e.
\begin{equation}
\label{ks1}
\left[ -\frac{1}{2} \nabla^2 + \pot{eff}^I(\br) \right] \phi^I_k(\br) = \varepsilon^I_k  \phi^I_k(\br),\mathrm{~with~}I=1,\ldots,N_S
\end{equation}
with the effective subsystem potential given by
\begin{equation}
\label{sks}
\pot{eff}^I (\br)=\underbrace{\pot{eN}^I (\br) + \pot{Coul}^I(\br) +\pot{xc}^I (\br)}_{\mathrm{same~as~regular~KS-DFT}} + \pot{emb}^I (\br).
\end{equation}
In FDE \cite{weso1993,weso2006}, $\pot{emb}$ appearing above is called embedding potential and is given by
\begin{align}
\label{emb}
\nonumber
\pot{emb}^I (\br)&=\sum_{J\neq I}^{N_S} \left[ \int \frac{\rhojrp}{|\br-\brp|} \d\brp - \sum_{\alpha\in J} \frac{Z_\alpha}{|\br-\bR_{\alpha}|} \right] + \\
&+\frac{\delta T_{\rm s}[\rho]}{\delta\rhor}-\frac{\delta T_{\rm s}[\rhoi]}{\delta\rhoir}+\frac{\delta E_{\rm xc}[\rho]}{\delta\rhor}-\frac{\delta E_{\rm xc}[\rhoi]}{\delta\rhoir}.
\end{align}

where $T_{\rm s}$, $E_{\rm xc}$ and $Z_\alpha$ are kinetic and exchange-correlation energy functionals, and the nuclear charge, respectively. At a first glance, the presence of the Coulomb potential in the above equation suggests that the method is not linear-scaling. However, two-electron integrals are never evaluated in our method as we employ Slater-type Orbitals for which no analytical two-electron integral formula is known.  Instead, an appropriate choice of fitting functions for the calculation of the Coulomb potential makes this integral easy to compute\cite{teve2001a}. Two additional simplifications are present in the FDE implementation: First, FDE uses subsystem-centered integration grids (significantly reducing the intgration time), and secondly, as the density of the non-active subsystems remain frozen during a subsystem's SCF procedure, their Coulomb potential is obtained once at the beginning of the SCF procedure and it is stored in memory. Ultimately, it is this that makes the method linear-scaling with respect to the number of subsystems\cite{jaco2008b}.

The density of the supersystem is found using \eqn{fde1} and \eqn{ks1}, for singlets $\rhor=2\sumi\sum_i^{{\rm occ}_I} \left| \phi^I_i(\br)\right|^2$. The above equations have not explicitly taken into account spin, however the full subsystem local spin density approximation equations can be retrieved elsewhere \cite{solo2012,good2011a}.

It has been shown \cite{pava2013a,solo2012,pava2011b} that FDE generates charge-localized states if one subsystem density is constrained to integrate to a number of electrons different from what would yield charge neutrality.  In this way, charge-localized broken symmetry states can be constructed by localizing an excess charge on a single molecular fragment (subsystem). \michele{This level of localization can also be achieved by the constrained DFT method of Van Voorhis \cite{kadu2012,wu2005,wu2006}. However, as opposed to the linear-scaling FDE method, constrained DFT scales identically to regular KS-DFT, i.e.\ as $N^3$.} FDE can construct an electron density featuring a hole on the donor (with corresponding ``initial'' wavefunction $\psi_D$) and one density where the hole is localized on the acceptor (with corresponding ``final'' wavefunction $\psi_A$). These states are not the solution of the full KS-DFT equations, thus a non-diagonal Hamiltonian matrix should be expected in the basis of $\psi_D$ and $\psi_A$. The off-diagonal elements of such Hamiltonian can be 
approximated by the following formula \cite{thom2009a}:
\begin{equation}
\label{cou}
 H_{DA}=\langle \psi_{D} | \hat H |\psi_{A}\rangle=S_{DA}E\left[\rho^{(DA)}(\br)\right].
\end{equation}
Here $\hat H$ is the molecular electronic Hamiltonian, and $\rho^{(DA)}(\br)$ is the transition density $\rho^{(DA)}(\br)=\langle \psi_D | \sum_{k=1}^{n_e} \delta (\br_k-\br) | \psi_A\rangle$, with $n_e$ being the total number of electrons in the system. The donor--acceptor overlap matrix elements are found by computing the following determinant:
\begin{equation}
\label{eq:s}
 S_{DA}=\mathrm{det}\left[\mathbf{S}^{(DA)}\right],
\end{equation}
where $\bS^{DA}_{kl}=\langle \phi_{k}^{(D)} | \phi_{l}^{(A)} \rangle$ is the transition overlap matrix and it is defined in terms of the occupied orbitals ($\phi_{k/l}^{(D/A)}$) making up the determinants $\psi_D$ and $\psi_A$ \cite{maye2002,thom2009a}
\begin{equation}
\label{Ivth}
 \rho^{(DA)} (\br)= \sum_{kl}^{\rm occ} \phi_{k}^{(D)} (\br)\left(\mathbf{S}^{(DA)}\right)_{kl}^{-1} \phi_{l}^{(A)}(\br).
\end{equation}
The Hamiltonian coupling is not $H_{DA}$, but it is generally reported as the coupling between the L\"{o}wdin orthogonalized $\psi_D$ and $\psi_A$. For only two states this takes the form,
\begin{equation}
\label{coulow}
 V_{DA} = \frac {1}{1-S_{DA}^2} \left(H_{DA}-S_{DA}\frac{H_{DD}+H_{AA}}{2}\right).
\end{equation}
For more information about the FDE formalism applied to charge transfer states, we refer the reader to Refs.\citenum{pava2011b,pava2013a}. 

\michele{In a previous work \cite{pava2013a}, it has been shown that the FDE method is comparable in accuracy with {\it ab-initio} methods for the calculation of charge transfer excitation energies for the test cases considered in that work. Generally speaking, semiempirical methods are extremely powerful and computationally cheap. However, we should mention that in a recent work\cite{adam2014} these methods, in particular DFT-based semiempiricals as the Density Functional tight binding (DFTB), produce an error of the electronic coupling around 40\% which reflects an overestimation of $\beta$ values by around 12\% (see for example Tables VI-XII in Adam et al.\cite{adam2014}). Regarding the computational cost, we mention that FDE scales linearly with the number of subsystems, which for a full-electron electronic structure method is a great advantage}.

\section{Computational details}
All calculations (effective electronic couplings, excitation energies, and site energies) were performed with a modified version of the Amsterdam Density Functional ({\sc Adf}) package \cite{teve2001a,jaco2008b}. GGA semilocal functionals, PW91 and PW91k, were used throughout for the exchange--correlation \cite{perd1991} and kinetic energy \cite{lemb1994} functionals, respectively. This combination of functionals (xc and kinetic parts) is known to reproduce $\pi$-stacking interaction energies in a subsystem DFT environment \cite{weso1997,weso1998,weso2003,gotz2009}. We used the TZP Slater--Type Orbital basis set throughout this work, except when noted.

The molecular systems chosen for the simulations are the following dimer combinations: DNA nucleobase (adenine, guanine and thymine)--DNA, DNA--Aminoacid (histidine, tyrosine, tryptophan), Aminoacid--Aminoacid and Aromatic (benzene, anthracene, indole, porphyrin, phenyl-porphyrin)--Aromatic. Finally, all DNA oligomer structures were obtained with the NAB program from {\sc AMBERtools} package \cite{amber} in the standard B-form.

\section{Results and Discussion}
\subsection{Distance dependence for hole-tunneling through vacuum}
\label{vac}
\begin{figure}[htp]
\begin{center}
\includegraphics[width=0.6\textwidth]{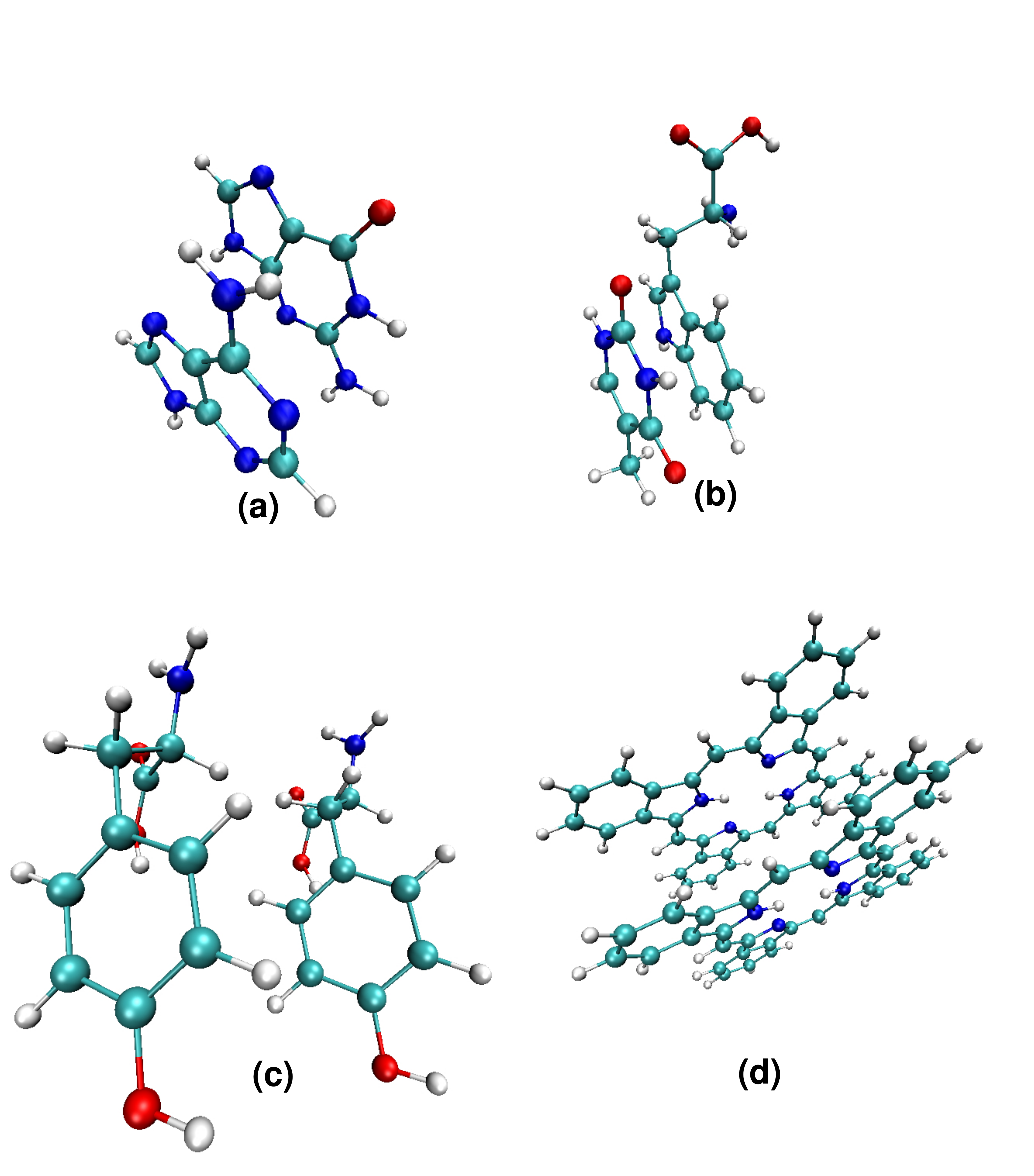} 
\end{center}
\caption{Sturctures of four of the 23 donor--acceptor dyads considered. (a) Adenine-Guanine, AG (DNA-DNA), (b) Thymine-Tryptophan T-Trp (DNA-aa), (c) Tyrosine-Tyrosine Tyr-Tyr (aa-aa) and (d) Tetrabenzoporphyrin dimer (Ar-Ar).}
\label{Forsys}
\end{figure}

In this section we present calculations of the coupling matrix element ($V_{DA}$) of hole transfer from a donor to an acceptor molecule through the vacuum. 
This means that the initial state of hole transfer is the donor molecule ($D$), and the final state the acceptor molecule ($A$), and no intermediate bridge states are considered. For this purpose, we chose 23 biologically relevant $\pi$-stacks dyads, and \ref{Forsys} shows a representative subset of them. In order to analyze the distance dependence of the coupling, donor-acceptor separations of 3-20 \AA\ are considered and 276 coupling calculations were ran in total.   

\begin{figure}[htp]
\begin{center}
\includegraphics[width=1.0\textwidth]{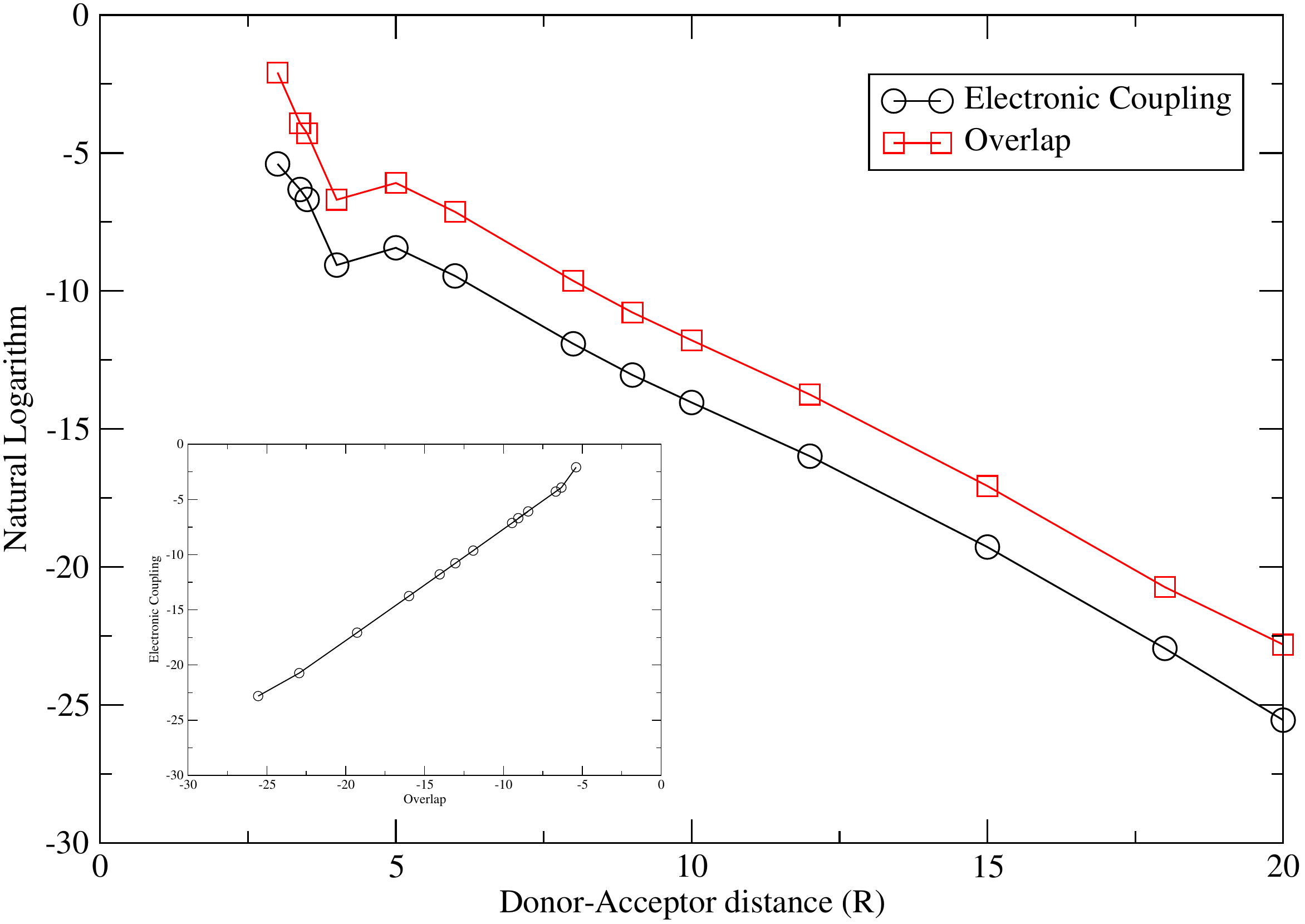} 
\end{center}
\caption{Distance dependence of coupling (black) and overlap (red) for the Adenine-Guanine (AG) system. Inset: $V_{DA}$ vs. $S_{DA}$ in a logarithmic scale. Color online. }
\label{ag}
\end{figure}

In Tables S1--S4 of the supplementary information, we collect the results for the donor--acceptor overlap, coupling, and site energies at variable donor--acceptor separations for the 23 dyads. The results show that at long ranges our couplings have a linear relation with the diabatic overlaps (see \ref{ag}). This behavior is expected, as it is reported in the literature\cite{migliore2011a,troisi2002,MRS1952}, where the electonic coupling and the overlap are found to be proportional to each other and the $\pi$-electron binding energy enters the proportionality constant. 

The FDE calculations were carried out similarly to Refs.\ \citenum{pava2013a,pava2011b}. In short, first the density of the two charge-localized states (either a hole on the donor or on the acceptor) are obtained by self-consistent FDE calculations (i.e.\ freeze--and--thaw\cite{jaco2008b}). Secondly, the total energy and Hamiltonian coupling among the resulting diabatic states are calculated in a post-SCF analysis according to \eqn{coulow}. The calculated couplings and the excitation energies (obtained solving the $2\times 2$ eigenvalue problem involving the Hamiltonian and the overlap matrix elements) of DNA nucleobases are compared against benchmark calculations \cite{voityuk2006a} and calculations carried out with the previous version of the code \cite{pava2013a}, see \ref{tabcom}.

\michele{The purpose of the set of donor--acceptor calculations is to test the numerical stability of our method for calculating long-range couplings which are at the base of coupling calculations in biosystems. In carrying out these initial tests, we identified two weaknesses of the previous implementation of our method which we have now completely cured. One was related to the density fitting routine, and one related to the overlap matrix inversion of \eqn{Ivth}. 

Let us first discuss the new density fitting routine. We have implemented in our code a new density fitting method \cite{fran2014} for evaluating the Coulomb potential associated with the transition density in \eqn{Ivth}, we now use a spline-based fit for the radial part of the density, and sets of atom-centered spherical harmonics for the angular dependence (ZLM fit, hereafter). ZLM fit offers the possibility to calculate the Coulomb potential to very high precision \cite{delley1990, fran2014}.}
\begin{table}[htp]
\begin{center}
\begin{tabular}{ccccccc}
\toprule
System & $V_{A/D}$ & $V_{A/D}^a$ &  $V_{A/D}^b$ & $E_{ex}$ & $E_{ex}^a$ & $E_{ex}^b$ \\
\hline
AA & 0.111 &0.092 & 0.004 & 0.234 &0.198 &0.097 \\
AG & 0.020 &0.177 & 0.044 & 0.235 &0.421 & 0.340 \\
GA & 0.052 &0.058 & 0.036 & 0.524 &0.530 & 0.560 \\
GG & 0.044 &0.051 & 0.051 & 0.403 &0.405 & 0.392 \\
GT & 0.102 &0.104 & 0.081 & 1.077 &1.082 & 1.175 \\
TG & 0.054 &0.056 & 0.061 & 0.654 &0.657 & 0.797 \\
TT & 0.096 &0.099 & N/A & 0.202 &0.208 \\
\bottomrule
\end{tabular}
\end{center}
\begin{flushleft}
\begin{center}
 $a$ FDE/PW91/TZP from Ref.\citenum{pava2013a}\\
$b$ CASPT2/6-31G* from Ref.\citenum{voityuk2006a}
\end{center}
\end{flushleft}
\caption{Hole transfer excitations and couplings for $\pi$-stacked DNA nucleobases at 3.38 \AA$^{-1}$ separation. All values in eV.}
\label{tabcom}
\end{table}
\michele{The results summarized in \ref{tabcom} are in generally in good agreement with both the CASPT2 benchmark and the prior calculations. In most cases, the electronic couplings calculated in this work are closer to the CASPT2 values than the results given in the previous work, hinting that the new density fitting routine is more accurate than the previous Slater-type orbital-based fit \cite{teve2001a} (STO fit, hereafter). The STO fit cannot be systematically improved as linear dependencies among the fit STO functions arize \cite{fran2014}. Conclusive evidence is provided by \ref{harfit}, where the relative error in the Coulomb energy of both diagonal and non-diagonal elements due to density fitting is shown. The values that we obtain with the ZLM fit method when a fine radial grid and high atomic angular momentum functions are used are several orders of magnitude smaller than with the STO fit, indicating that the new fit is accurate and that numerically stable excitation energies and couplings are obtained. We also noticed (not shown) that the coupling and excitation energies calculated with ZLM fits of varying accuracy are almost identical, indicating that the ZLM fit provides a balanced fit of the diabatic densities and of the transition density.}
\begin{table}[htp]
\begin{center}
\begin{tabular}{ccccccc}
\toprule
\multirow{2}{*}{System} & \multicolumn{2}{c}{\multirow{2}{*}{STO fit (Ref.\citenum{teve2001a})}} & \multicolumn{4}{c}{ZLM fit (Ref.\citenum{fran2014})} \\ 
       &  & & \multicolumn{2}{c}{L=4, G=3} & \multicolumn{2}{c}{L=12, G=14} \\ 
\hline
& $\sigma_{AA/DD}$ & $\sigma_{AD}$ & $\sigma_{AA/DD}$ & $\sigma_{AD}$ & $\sigma_{AA/DD}$ & $\sigma_{AD}$ \\
\hline
AA &  16.781 &  3.476 &  4.087 &  6.048 &  0.038 &  0.050 \\
AG &  12.698 & 12.605 &  4.260 &  4.204 &  0.043 &  0.043 \\
GA &  14.122 & 13.857 &  5.443 &  4.974 &  0.041 &  0.052 \\
GG &  10.133 & 13.579 &  5.015 &  4.139 &  0.036 &  0.000 \\
GT &  12.555 & 18.026 &  6.208 &  5.893 &  0.059 &  0.072 \\
TG &   8.968 &  6.488 &  4.859 &  4.930 &  0.067 &  0.058 \\
TT &  13.379 & 12.376 &  3.749 &  4.378 &  0.037 &  0.048 \\
\bottomrule
\end{tabular}
\end{center}
\caption{Coulomb relative error in part per thousand (i.e.\ fit error multiplied by 10$^3$ and divided by the total Coulomb energy) for the two different fitting routines employed. The fit error is denoted with $\sigma_{AA/DD}$ and $\sigma_{AD}$ for the diagonal and the off-diagonal Hamiltonian matrix elements, respectively. L is the maximum angular momentum used in the angular part of the ZLM fit, and G is a measure of the radial grid density. For example, G=5 is a five-times denser radial grid than G=1.}
\label{harfit}
\end{table}

We will now discuss the long-range behavior of the calculated couplings and the numerical inaccuracies arising with the inversion of the transition overlap matrix in \eqn{Ivth}. The Hamiltonian matrix element formula in \eqn{cou} holds for non-orthogonal states. When the wavefunctions of the two diabatic states are represented in terms of Slater determinants, the non-orthogonality condition is translated to a non-singularity condition for the transition overlap matrix. The transition overlap matrix might become singular if one or more orbitals of one diabatic state are orthogonal to all the occupied orbitals of the other diabatic state. This can happen in two distinct cases: (1) orthogonality by symmetry considerations, and (2) spatial separation of the orbitals. Case (1) almost never occurs, as all the geometries considered are not exactly symmetric. Here we say ``almost'' as we seldomly encountered such symmetry related orthogonalities. Case (2), instead, will always occur when donor and acceptor moieties are separated by a distance large enough that the maximum overlap of the atomic orbitals employed becomes zero. In both cases, the Hamiltonian matrix element formula in \eqn{cou} fails. 
 
A workaround which was proposed in a previous work \cite{pava2013a} prescribed the inversion of the transition overlap matrix only in the orbital subspace where the matrix is non-singular. Numerically, however, the stability of this algorithm is dictated by the relative magnitude of the overlap matrix elements. In this work, first a singular value decomposition of the overlap matrix is carried out and then if one of the singular values is lower than a threshold set by the user, the approximate Penrose inversion is adopted \cite{pava2013a}. An inversion threshold of $10^{-3}$ was set as default. This protocol was implemented in the {\sc Adf} program as a feature of the {\sc electrontransfer} keyword.

For the DNA systems, the default inversion threshold was appropriate in most cases. However three systems stand out: AG, GA and TT nucleobase pairs for which a threshold of $10^{-2}$ was adopted to avoid numerical inaccuracy in the inversion. The mentioned systems for some donor-acceptor distance, specifically 4.0 \AA\ for AG, 3.5 \AA\ and 8.0 \AA\ for GA and 9.0 \AA\ for TT, feature near singularity of the overlap matrix due to symmetry considerations (case 1 above). This can be seen from the value of the $S_{DA}$ overlap decreasing only in those mentioned distances (see Tables S1--S4 in the supplementary materials). \ref{ag} depicts the natural logarithm of the coupling plotted against donor-acceptor distance for AG, the system for which the most erratic behavior of the coupling due to symmetry-related quasi orthogonality was observed. From the plot it is clear that the overlap element and the coupling are linearly related (see also the inset of \ref{ag}) also at short range. This 
behavior is basis set independent, as it is also recovered when employing a larger basis set (TZ2P) see Table S4. This basis set contains an additional shell of polarization functions. It is substantially larger than the TZP set. The couplings calculated with the TZ2P basis set display the same trend as the TZP ones. Also, in Table S4 we show that in the worst cases the effective couplings and overlaps differ from the ones reported with TZP basis by a dismal 0.2 meV and 0.02, respectively. Thus, the presented couplings are robust and basis set independent. 

\begin{table}[htp]
\begin{center}
\begin{tabular}{ccc}
\toprule
System & $\beta$ (\AA$^{-1}$) & $\sigma_\beta$ \\
\hline
DNA &&\\
\hline
GG & 2.38 & 0.06 \\
AA & 2.63 & 0.11 \\
AG & 2.21 & 0.09 \\
TG & 2.46 & 0.10 \\
GA & 2.46 & 0.12 \\
GT & 2.44 & 0.03 \\
TT & 2.55 & 0.14 \\
\hline
Aminoacid-DNA & & \\
\hline
His-A & 2.52 & 0.03 \\ 
Trp-A & 2.72 & 0.05 \\
Tyr-A & 2.51 & 0.15 \\
His-G & 2.68 & 0.05 \\
Trp-G & 2.51 & 0.10 \\
Tyr-G & 2.49 & 0.04 \\
His-T & 2.71 & 0.10 \\ 
Trp-T & 2.42 & 0.06 \\
Tyr-T & 2.26 & 0.07 \\
\hline 
Aromatics && \\ 
\hline
Anthracene & 2.65 & 0.03 \\
Benzene & 2.46 & 0.05 \\
Tetrabenzoporphyrin & 2.71 & 0.03 \\
Indole & 2.39 & 0.04 \\
Porphyrin & 2.32 & 0.03 \\
Trp-Trp & 2.23 & 0.04 \\
Tyr-Tyr & 2.38 & 0.04 \\
\bottomrule
\end{tabular}
\end{center}
\caption{Tunneling decay factors and standard deviation in the fitted $\beta$ values for the DNA nucleobase pairs, aminoacid-nucleobase, and aromatic dyads.}
\label{tabbeta}
\end{table}

The long range behavior of the couplings was characterized by a fit against the exponential decay law in \eqn{j-m-l}. 
The decay parameter, $\beta$, is defined, according to \eqn{marcus}, in terms of a fit of the distance dependence of the coupling to the following exponential function:
\begin{equation}
\label{V_vlm}
 V_{DA} (R) = V_0 \exp\left(-\frac{\beta}{2} R\right).
\end{equation}
By taking the natural logarithm, \eqn{V_vlm} takes a linear form:
\begin{equation}
 \label{j-m-l}
 \ln\lvert V_{DA}^2 (R) \rvert= \ln\lvert V_0^2 \rvert - \beta R.
\end{equation}
As we can see in the supplementary information {Figures S1--S4} and in \ref{tabbeta}, the electronic couplings decay exponentially as expected \cite{gies2002}, and the fits to the exponential function deliver excellent standard deviations for the fit parameters. 

Turning to the excitation energies in Tables S1--S4, we notice that they are within the thermally accessible range of the donor-acceptor ionization potential difference. For instance, let us consider the AG and AA systems. In Table S1, these two systems display a similar excitation energy. This can be explained by considering that the ionization potentials of the involved nucleobases are similar (IP$_{\rm A}\simeq 8.0$ eV and IP$_{\rm G}\simeq 7.8$ eV \cite{mish2009}), with a difference of only about $\sim 0.2$ eV indicating that the excitation energies expected at room temperature (considering a maximum of 0.2 eV fluctuation \cite{voityuk2007a,pava2010a}) should range between 0-0.4 eV for AG and  0--0.2 eV for AA. 

Despite of the good fit to the exponential decay law, the calculated decay factors are not comparable to the experimentally determined ones ($\beta\sim 0.7-1.5$ for DNA nucleobases \cite{gies2004}, and $\beta\sim 0.8-1.2$ for aminoacids \cite{wink2000}). Such an overestimation of the decay factors by our calculations is expected, as the hole in this first set of calculations tunnels from donor to acceptor through the vacuum. In the experiments, however, the tunneling barrier is always reduced from the vacuum level by the existence of other molecules populating the space in between donor and acceptor. These molecules are often called ``bridge'' and the resulting tunneling is termed bridge-mediated tunneling or superexchange. In the following sections, we will evaluate bridge-mediated superexchange couplings in two DNA oligomers employing the effective two-state approximation. However, the two-state approximation is bound to break down when near-resonant bridge states are present in the oligomer.

\subsection{All-electron superexchange couplings in DNA oligomers: Quantifying the environmental effects}
This section presents calculations of electronic couplings of guanine-to-guanine hole transfer in the \gtg\ DNA oligomer, with $N=$0-3. We will use a shorthand notation \sgtg, hereafter, for this system. By including in the computations increasingly complex molecular environments, the effect of the surrounding can be determined by analyzing the trend in the decay factor, $\beta$, when going from interacting nucleobases of a single strand to a dephosphorilated B-DNA that includes the counter strand and the sugar groups. 

\begin{figure}[ht]
\begin{center}
\includegraphics[width=0.8\textwidth]{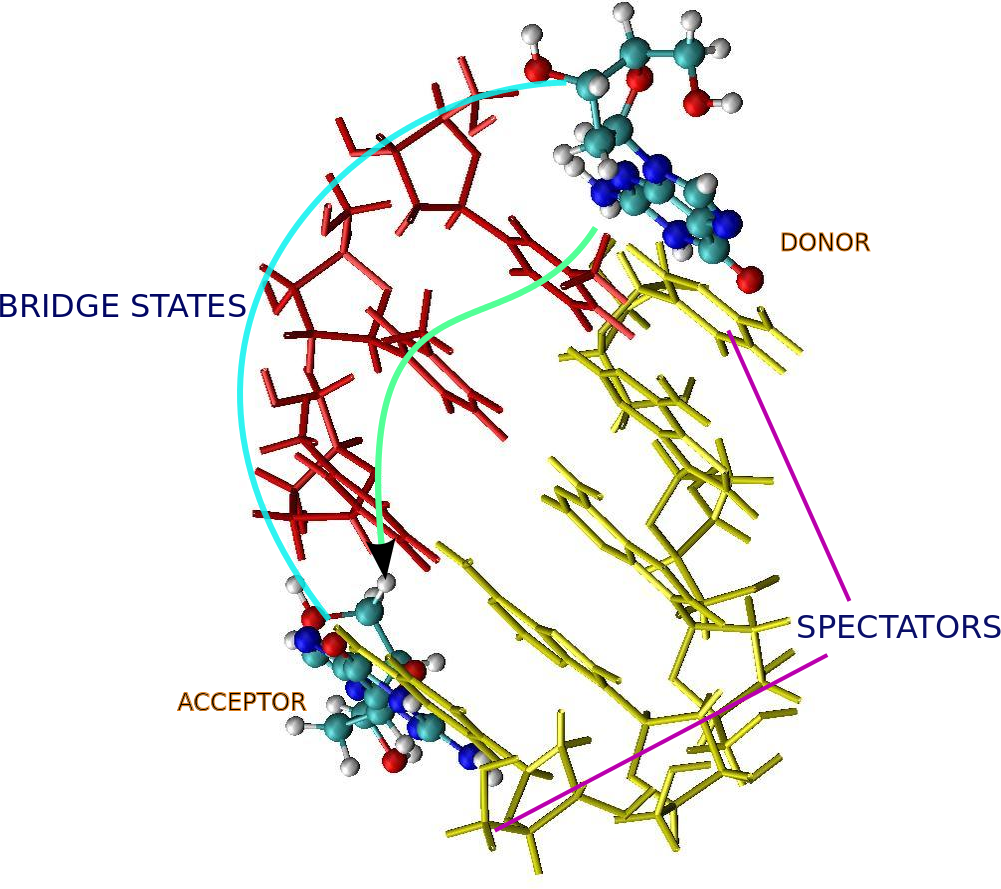} 
\end{center}
\caption{The dephosphorilated \sgtg\ B-DNA oligomer employed in the hole transfer coupling calculations. As the figure depicts, the hole tunnels from the bottom guanine (in balls and sticks) to the top guanine. The tunneling wall is provided by a series of three thymines (red branch, labeled as ``bridge''). The counterstrand, C(A)$_N$C, acts as a solvating environment (in yellow, labeled as ``spectators'') and no hole is allowed to localize on it.}
\label{orbOstrand}
\end{figure}

The starting point of the calculations is a completely dry B-DNA structure of \sgtg. The structures considered lack water molecules, metal counterions and phosphate linker groups. This is because the applicability of FDE is restricted to non-covalently bound molecular fragments. Consequently, appropriate modifications to the B-DNA structure had to be made: we have removed the phosphate groups and capped the dangling bonds with hydrogen atoms at 1.09 \AA\ from the bonding atom. The resulting structure of the modified \sgtg\ is depicted in \ref{orbOstrand}. It is important to point out that the calculations were carried out with a fixed nuclear structure, i.e.\ thermal dissipation, as well as vibronic effects were not contemplated at all in this work, and will be the focus of a future study. 

\subsubsection{Theory of effective donor--bridge--acceptor couplings}
The \atn\ bridge separating donor and acceptor assists the hole transfer by lowering the effective tunneling barrier. This effect can be taken into account by considering the full Hamiltonian and overlap matrices of the hole pseudoparticle defined in \eqs{cou}{eq:s}. Namely,
\begin{equation}
 \mathbf{H}=\left(
\begin{array}{crr}
 E_{D} & ... & H_{DA}\\
 \vdots &  \mathbf{H}_{B} & \vdots\\
 H_{AD} & ... & E_{A} 
\end{array}
\right)
 ,\ 
  \mathbf{S}=\left(
\begin{array}{crr}
  1 & ... & S_{DA}\\
 \vdots &  \mathbf{S}_{B} & \vdots\\
 S_{AD} & ... & 1
\end{array}
\right).
\end{equation}
In the matrices above, a clear distinction between matrix elements between states where the hole is localized on the bridge molecules or on the donor and acceptor molecule has been labeled. Taking into account that during the hole tunneling the bridge states are virtually occupied by the hole, an effective coupling can be obtained by reducing the generalized eigenvalue problem constructed with the above Hamiltonian and overlap matrices to a 2$\times$2 effective eigenvalue problem \cite{Nitzan_book,mcweeny,skou1993}. A L\"{o}wdin orthogonalization of the basis set yields a transformed Hamiltonian matrix, $\mathbf{\tilde V}$, from which the following bridge-mediated effective hole coupling is derived \cite{even1992,Nitzan_book,newton1991,lowd1963,Larsson1981,Priya1996}:
\begin{equation}
\label{efcou}
 V_{DA}(E) = {\tilde V}_{DA} + \underbrace{\mathbf{\tilde V}_{DB}^{T} \mathbf{G}_{B}(E)\mathbf{\tilde V}_{BA}}_{V_{\rm bridge}},
\end{equation}
where the superscript $T$ stands for transpose, $\mathbf{G}_{B}(E)$ is the Green's operator, defined as
\begin{equation}
\label{green}
 \mathbf{G}_{B}(E) = -( \mathbf{\tilde V}_{B}-E~\mathbf{\tilde I}_B)^{-1},
\end{equation}
and $\mathbf{\tilde V}_{DB/BA}$ is the row vector of the transformed Hamiltonian collecting the couplings between the donor/acceptor with the bridge states. Generally, $E$ appearing above is the energy at which the tunneling event occurs (i.e.\ at the crossing seam of the Marcus parabolas). As we only considered static geometries of the DNA oligomers, the Hamiltonian eigenvalues corresponding to the donor-acceptor energies do not coincide. With that, the tunneling energy, $E$, is not well defined \cite{marc1987}. \michele{A natural choice of $E$ is to place it between $E_{A}$ and $E_B$, with a common choice being $\frac{E_D+E_A}{2}$. For example, this choice is invoked by several works in the literature \cite{hatc2008,newton1991,marc1987,voityuk2012}. Others\cite{voit2002}, have favored the choice $E=E_D$, which is non-symmetric (i.e.\ forward CT become not equivalent to backward CT), however, it is still a valid choice. It was Beratan \cite{Bera1984} and Marcus \cite{marc1987} who showed early on the weak dependence of the coupling with 
respect to the choice of tunneling energy within $E_D$ and $E_A$. Thus, we adopt $E=\frac{E_D+E_A}{2}$ in all calculations.}

In the following section, we will use \eqn{efcou} to estimate the hole superexchange coupling matrix element in DNA oligomers.  It is convenient to consider the two components of the coupling separately: The component related to the tunneling through the vacuum [${\tilde V}_{DA}$ in \eqn{efcou} also called ``through space'' coupling], and the superexchange contribution from the bridge states ($V_{\rm bridge}$). This distinction offers insight to the factors influencing the overall superexchange coupling. However, it is known that \eqn{efcou} is valid only when the donor, bridge and acceptor states are weakly mixed. \michele{Hatcher et al.\ \cite{hatc2008} found that if A is present between two Gs, there is a high chance that dynamical fluctuations will lead to a situation where the energy levels of A are in resonance with G, thus, undermining the two-state model and the validity of \eqn{efcou} (formally valid only for non-resonant tunneling through the bridge)}. Non-resonant tunneling is indeed the case for G(T)$_N$G.
 For G(A)$_N$G, near degeneracies of the donor/acceptor--bridge system can arise (see subsection 4.2 and figure 5). For the particular conformations that we adopt in this study, we indeed encountered near-degeneracies for G(A)$_2$G and G(A)$_3$G oligomers.

\subsubsection{Decay rate of the hole coupling in single and double stranded \sgtg\ oligomer}
In this section, we present and analyze the distance decay rates as well as the effect of the molecular environment on the effective hole transfer coupling $V_{DA}$ for four \sgtg\ model systems: single/double strand, and with/without ribose groups. When the ribose groups are included in the calculations, they are attached covalently to the nucleobases. This effectively increases the size (in terms of number of atoms) of all the participating charge transfer states. The largest system considered is the double strand with ribose groups and counts 308 atoms and 1322 electrons. A summary of the coupling values is given in \ref{tabCM}.

\begin{table}[htp]
\begin{center}
\begin{tabular}{lrrrr}
\toprule
 & ${\tilde V}_{DA}$ (meV) & $V_{\rm bridge}$ (meV) & E$_{\rm DB}$ (eV) & E$_{\rm BA}$ (eV) \\
\hline
\multicolumn{5}{c}{{\sc Single Strand no Ribose}} \\ 
\hline
GG        & 78.13   &      &       &     \\
GTG       & 0.76   & 12.46  & 0.71   & 0.50 \\
G(T)$_2$G & 0.01   &  1.13  & 0.79   & 0.66 \\
G(T)$_3$G & --  &  0.09  & 0.79   & 0.77 \\
\hline
\multicolumn{5}{c}{{\sc Double Strand no Ribose}} \\
\hline
GG        & 92.6   &      &        &     \\
GTG       & 0.65 & 7.66   & 0.93   & 0.96 \\
G(T)$_2$G & 0.01 & 0.47   & 1.11   & 0.94 \\
G(T)$_3$G & -- & 0.02   & 0.99   & 1.16 \\
\hline
\multicolumn{5}{c}{{\sc Single Strand with Ribose}} \\
\hline
GG        & 71.38   &      &        &     \\
GTG       & 0.18 & 25.01  & 0.43   & 0.37 \\
G(T)$_2$G & 0.02 &  1.70  & 0.58   & 0.37 \\
G(T)$_3$G & -- &  0.21  & 0.41   & 0.41 \\ 
\hline
\multicolumn{5}{c}{{\sc Double Strand with Ribose}} \\
\hline
GG        & 91.07   &      &        &     \\
GTG       & 0.02 & 7.35   & 0.62   & 0.87 \\
G(T)$_2$G & 0.02 & 0.61   & 0.93   & 0.60 \\
G(T)$_3$G & -- & 0.02   & 0.50   & 0.82 \\
\bottomrule
\end{tabular}
\end{center}
\caption{Through-space and through-bridge electronic coplings and tunneling energy gaps for single and double strand \sgtg\ B-DNA, including the effects of the backbone (sugars). A -- is shown for values below 0.01meV.} 
\label{tabCM}
\end{table}

It is interesting to notice that the two components to the overall coupling, $\tilde V_{DA}$ and $V_{\rm bridge}$, follow an inverted trend. 
Let us first analyze the embedding effects due to the ribose groups on the direct $\tilde V_{DA}$ couplings. It is arguable that the presence of any environment induces some slight charge delocalization in the adiabatic states simply because additional quantum states become accessible. However, we see that the ribose groups have the effect of localizing further the hole's wavefunction to the nucleobases in turn decreasing the direct nearest-neighbor coupling. We notice such trend for all the \sgtg\ oligomers. The non-nearest-neighbor couplings are only marginally affected (differences of a few tenths of meV).
 
The picture is inverted for the superexchange component of the coupling, $V_{\rm bridge}$. There, we notice that the ribose groups increase this coupling term consistently. We rationalize these contrasting trends by generalizing that the through-space effect is due to a further localization of the hole's wavefunction, while the through-bridge effect is due to an uneven solvation by the ribose of the G and T nucleobases and due to the H-bonding interaction with the partner strand. We remind the reader that the effects related to the phosphates groups are not contemplated here, as these groups were not included in the structures used for coupling and site energy calculations. A closer inspection of the data in \ref{tabCM} reveals that the double strand environmental effects overpower completely the effects due to the ribose groups. This is an important finding, as it indicates that the wavefunction and energetics of the hole in DNA is almost completely determined by the interactions with the counterstrand rather than by the interactions with the covalently bound ribose groups. We will further analyze this aspect later on when considering the decay factors, $\beta$.

\begin{figure}[!t]
\begin{center}
\includegraphics[width=1.0\textwidth]{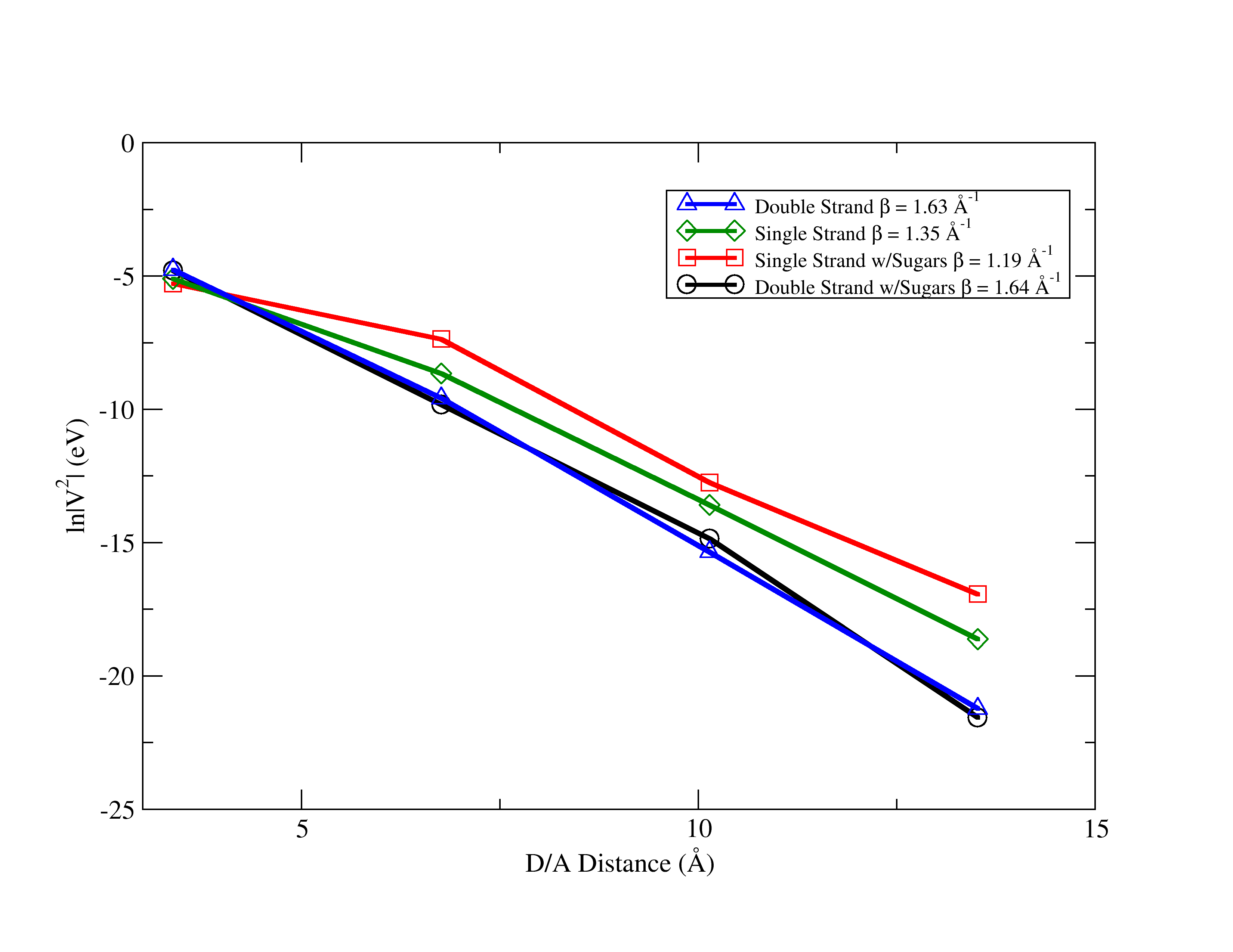}
\end{center}
\caption{Distance dependence of the bridge mediated tunneling superexchange $\ln|V_{DA}|$ for \sgtg\ with $N=$0-3}
\label{gtgbeta}
\end{figure}

\begin{figure}[htp]
\begin{center}
\includegraphics[width=1.0\textwidth]{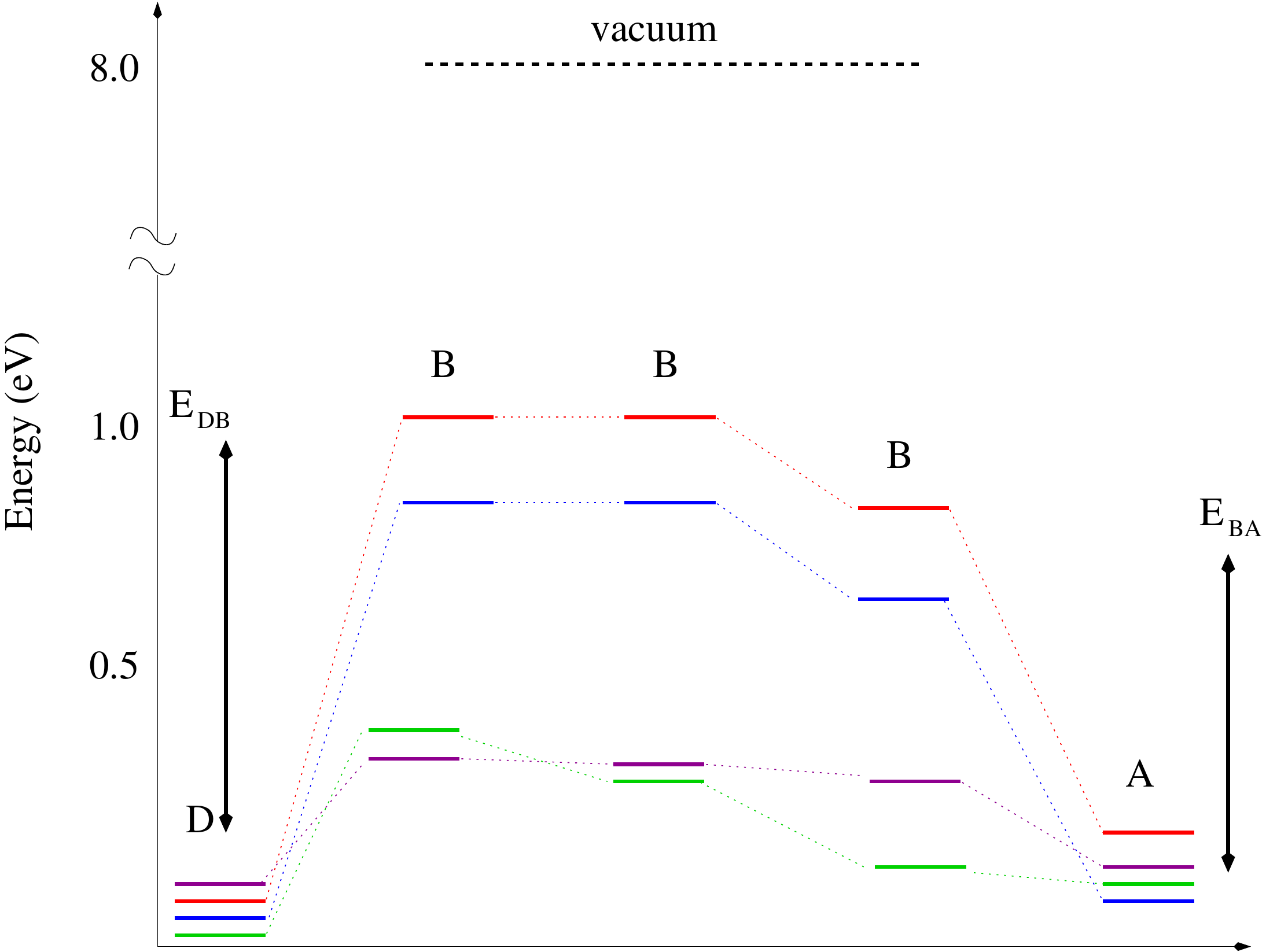}
\end{center}
\caption{Energy level diagram for \sgtg\ with the single strand (in blue), and the double strand (in red). While for \sgag , the single strand is in violet and the double strand in green. The donor's ionization potential (guanine in 5$^\prime$ position) was taken as the zero of energy.}
\label{etun}
\end{figure}

The full picture of the environmental effects on the hole-transfer couplings is captured by \ref{gtgbeta}. There, the logarithmic decay of the effective couplings with the donor--acceptor distance is plotted. The trends are explained by inspecting the relative energy levels of donor/acceptor and the bridge states in the single strand and in the double strand oligomers presented in \ref{etun}.
The figure shows that the energy levels of the bridge states change in going from the 5$^\prime$ to the 3$^\prime$ side of the DNA oligomer. This behavior has been characterized before both experimentally \cite{sait1998} and theoretically \cite{voit2000} (although using semiempirical methods). Our calculations reproduce the experimental trend that the nucleobases in 3$^\prime$ (right hand side in \ref{etun}) have lower ionization potential than the same nucleobases placed at the 5$^\prime$ position \cite{sait1998}.
Considering the average tunneling barrier height, calculated by taking the average of all energy differences between each bridge state with D/A states (see \ref{tabCM}), a dependency of the electronic coupling with the barrier height is evident. The taller the barrier, the larger the $\beta$. This result is in line with the concept that the square of the coupling is related to the transmission probability of tunneling through the wall of potential energy separating the donor and the acceptor \cite{Nitzan_book}.

To further characterize the solvating effect of the DNA backbone on the electronic couplings in \sgtg , we included the deoxyribose groups on all the nucleobases present in the model system (including the spectator counterstrand). As before, we analyzed the differences in the hole transfer parameters for the single and double stranded cases. \ref{gtgbeta} and \ref{tabCM} show that the presence of the sugar groups does not change the picture significantly and we recover a similar trend of couplings and site energies as for the structures without sugars. However, from \ref{tbeta} we notice that while the couplings in the single stranded \sgtg\ are affected by the presence of the sugar groups ($\beta$ values of 1.34 to 1.19, respectively), the double stranded system is not appreciably affected by the sugar groups. Our calculations show that while the site energies are generally lowered by the presence of the sugars, the stabilizing effect of the counterstrand and the effect of the sugar groups on the localization of the hole wavefunction have opposite effects on the coupling. This results in a small variation of the calculated couplings when the sugar groups are included. The calculated $\beta$ values are in broad agreement with the accepted values for DNA, however overestimate the measurements of Giese et al.\ \cite{gies2004} ($\beta=0.6$).
\begin{table}[htp]
\begin{center}
\begin{tabular}{lrr}
\toprule
& $\beta$ & $\sigma_\beta$ \\
\hline
\multicolumn{3}{c}{{\sc \sgtg\ Sequence }} \\ 
\hline
Single Strand  & 1.34                   &  0.08 \\
Double Strand & 1.63                  & 0.05 \\
Single Strand with Sugars & 1.19  & 0.14 \\
Double Strand with Sugars & 1.64  & 0.09 \\
\hline
\multicolumn{3}{c}{{\sc \sgag\ Sequence}} \\
\hline
Single Strand & 1.01         & 0.13\\
Double Strand & 1.56    &  0.33 \\
\bottomrule
\end{tabular}
\end{center}
\caption{Tunneling decay factors and Asymptotic Standard Error for all \sgtg\ B-DNA sequences considered.} 
\label{tbeta}
\end{table}

\begin{table}[htp]
\begin{center}
\begin{tabular}{lrrrr}
\toprule
 & ${\tilde V}_{DA}$ (meV) & $V_{\rm bridge}$ (meV) & E$_{\rm DB}$ (eV) & E$_{\rm BA}$ (eV) \\
\hline
\multicolumn{5}{c}{{\sc Single Strand no Ribose}} \\ 
\hline
GG        & 78.13   &      &       &     \\
GAG       & 0.67  &  5.43 & 0.32  & 0.17 \\
G(A)$_2$G & 0.01  & 1.43  & 0.37  & 0.15 \\
G(A)$_3$G & --  & 0.44 & 0.22  & 0.29 \\
\hline
\multicolumn{5}{c}{{\sc Double Strand no Ribose}} \\
\hline
GG        & 92.6   &      &        &     \\
GAG       & 0.69    & 0.78   & 0.35 & 0.25\\
G(A)$_2$G & 0.01    & 0.07  & 0.32 & 0.24 \\
G(A)$_3$G & -- & 0.04 & 0.27 & 0.28 \\
\bottomrule
\end{tabular}
\end{center}
\caption{Electronic couplings and tunneling wall heights of the \sgag\ systems. A -- is shown for values below 0.01 meV.} 
\label{tab_ade}
\end{table}
\begin{figure}[htp]
\begin{center}
\includegraphics[width=1.0\textwidth]{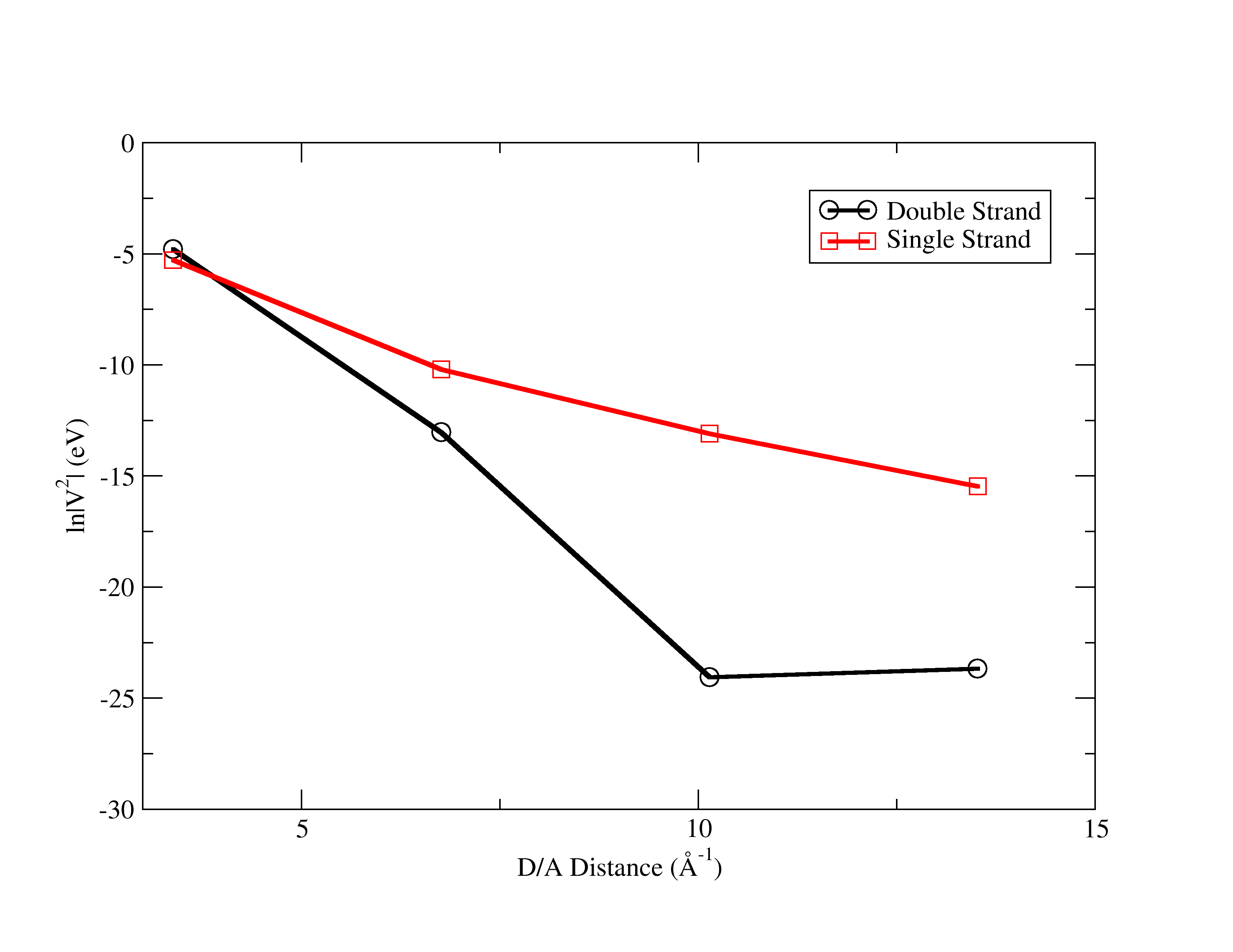}
\end{center}
\caption{Distance dependence of the bridge mediated tunneling superexchange $\ln|V_{DA}|$ for \sgag\ with $N=$0-3}
\label{allade}
\end{figure}

In this work, we also analyzed the system \sgag\ with $N=$0-3, for single and double stranded DNA. Although, as we mentioned above the DNA oligomers with bridging adenines invalidate the two-state approximation, when we apply \eqn{efcou} we assume a non-resonant behavior between guanines and adenines which is an approximation.  Borrowing from the \sgtg\ system that the effect of the sugars is not significant, for this final test we omitted the calculations with the ribose groups. All calculations were carried out in the same way as the \sgtg\ systems, however, we used the transition overlap inversion threshold of $10^{-2}$ for the simulations involving $AG$ and $GA$ pairs as we had determined these thresholds in the corresponding calculations carried out in the vacuum (Section \ref{vac}). \ref{tab_ade} shows that, once again, the couplings follow an exponential decay rate when the distance is increased, and similar to \sgtg , single strands sequences have lower tunneling wall height which results in a lower $\beta$ (see \ref{allade}). The $\beta$ values of 1.01 \AA$^{-1}$ and 1.57 \AA$^{-1}$ (see \ref{tbeta}) are in agreement with experimental data \cite{gies2002,gies2000}.

Turning to the energy levels, we see on \ref{etun} that the second strand system features bridge energy values decaying much faster than the single strand ones. The general trend of the site energies for the bridge is in agreement with previous theoretical predictions \cite{voit2000}, however, the near degeneracy experienced by the 3$^\prime$-most adenine could not be predicted and it is likely the result of the specific nuclear geometry chosen in this work. \michele{We believe this near-degeneracy is the reason for the fact that the coupling does not follow an exact exponential decay law in \ref{allade}. This is problematic and may undermine the applicability of \eqn{efcou} as previously discussed by Hatcher et al.\cite{hatc2008}.}

We also notice that the uneven stabilization of the bridge states and donor/acceptor states is much more pronounced in the \sgtg\ system than in the \sgag\ system. After inspection of the overall electrostatics of the interaction between G:C and T:A \cite{mill1990}, we notice that T has a strong permanent dipole pointing towards A, similarly to C:G. Instead, A has a much weaker dipole compared to C or T and thus upon contact of the GTG strand with the CAC strand the cytosines will stabilize much more the holes on Gs than the adenines can stabilize the holes on Ts, hence the tunneling wall increases from single strand to double strand. for the \sgag\ system this effect is much dampened by the fact that donor, acceptor, and bridge states are similarly solvated by the counterstrand.
 
\section{Conclusions}
Characterizing the effects of the molecular environment on the through space and through bridge hole transfer couplings in biological systems is currently out of reach of standard all-electron electronic structure methods due to the large system sizes needed in the simulations. Subsystem DFT offers a way to include the environmental effects from first principles with no need to parametrize the interactions between subsystems. 

In this work, we have showed that the Frozen Density Embedding formulation of subsystem DFT is capable of tackling biosystems of realistic sizes, such as double stranded DNA pentamers. The simulations focused on the hole transfer couplings of 23 dyads relevant to biology as well as two DNA oligomers, \sgtg\ and \sgag. While the calculations on the dyads were carried out for benchmark purposes, and are fairly standard not revealing any unexpected results. The calculations on the DNA oligomers, instead, uncovered new paradigms related to the interactions of the hole in DNA with its molecular environment.

Our calculations of hole transfer in DNA reproduce experimental findings regarding the preference of the hole for the 3$^\prime$ position rather than the 5$^\prime$. In agreement with simple arguments based on electrostatics, our calculations show that the bridge states (thymines) in \sgtg\ experience a dramatically different environmental effect than the donor and acceptor (guanines). Again, in accordance with the simple electrostatic picture, the same effect is not noticed in the \sgag\ system. 

When the through space and through bridge couplings are inspected, our calculations show that the effects of the ribose groups and the nucleobases in the counterstrand are opposite and different in magnitude depending on the oligomer size. We conclude, however, that the effect of the counterstrand completely overpowers any effect due to the presence of the ribose groups.
Our calculated decay factors ($\beta$) feature excellent standard deviation, and satisfactory agreement with the experimentally determined ones. 

The major limitation of our calculations rests in the absence of nuclear dynamics. As dynamics plays a major role in modulating the couplings and energies in biological hole transfer, we commit to investigate such dynamical effects in a follow up work.

\section*{Acknowledgements}
We thank the NSF-XSEDE program (award TG-CHE120105) for computational resources, and the office of the Dean of FASN of Rutgers-Newark for startup funds. We also thank Agostino Migliore and Alexander Voityuk for enlightening discussions.

\section*{Supplementary Information}
This information is available free of charge via the Internet at http://pubs.acs.org/.

\providecommand*{\mcitethebibliography}{\thebibliography}
\csname @ifundefined\endcsname{endmcitethebibliography}
{\let\endmcitethebibliography\endthebibliography}{}

\end{document}